\documentclass[twocolumn,prb,superscriptaddress,longbibliography,citeautoscript]{revtex4-1}
\usepackage{amsmath,amssymb}
\usepackage{bm}
\usepackage{graphicx}
\usepackage{color} 
\usepackage{setspace}
\usepackage{subfigure}

\begin{document}
\title{Nonlinear localization of light using the Pancharatnam-Berry phase}

\author{Chandroth P. Jisha}
\email{Email: cpjisha@gmail.com}
\affiliation{Institute of Applied Physics, Abbe Center of Photonics, Friedrich-Schiller-Universit{\"a}t Jena, Albert-Einstein-Str. 15, 07745 Jena, Germany}
\author{Alessandro Alberucci}
\email{alessandro.alberucci@gmail.com}
\thanks{These authors contributed equally to this work.}
\affiliation{Institute of Applied Physics, Abbe Center of Photonics, Friedrich-Schiller-Universit{\"a}t Jena, Albert-Einstein-Str. 15, 07745 Jena, Germany}
\author{Jeroen Beeckman}
\affiliation{Ghent University, ELIS Department, Technologiepark-Zwijnaarde 15, 9052 Gent, Belgium}
\author{Stefan Nolte}
\affiliation{Institute of Applied Physics, Abbe Center of Photonics, Friedrich-Schiller-Universit{\"a}t Jena, Albert-Einstein-Str. 15, 07745 Jena, Germany}
\affiliation{Fraunhofer Institute for Applied Optics and Precision Engineering, Albert-Einstein-Stra{\ss}e 7, 07745 Jena, Germany}

\date{\today}


\maketitle
%

\textbf{Since its introduction by Sir Michael Berry in 1984, geometric phase became of fundamental importance in physics, with applications ranging from solid state physics to optics. In optics, Pancharatnam-Berry phase allows the tailoring of optical beams by a local control of their polarization. Here we discuss light propagation in the presence of an intensity-dependent local modulation of the Pancharatnam-Berry phase. The corresponding self-modulation of the wavefront counteracts the natural spreading due to diffraction, i.e., self-focusing takes place. No refractive index variation is associated with the self-focusing: the confinement is uniquely due to a nonlinear spin-orbit interaction. The phenomenon is investigated, both theoretically and experimentally, considering the reorientational nonlinearity in liquid crystals, where light is able to rotate the local optical axis through an intensity-dependent optical torque. Our discoveries pave the way to the investigation of a new family of nonlinear waves featuring a strong interaction between the spin and the orbital degrees of freedom.
}

\begin{figure}
\includegraphics[width=0.49\textwidth]{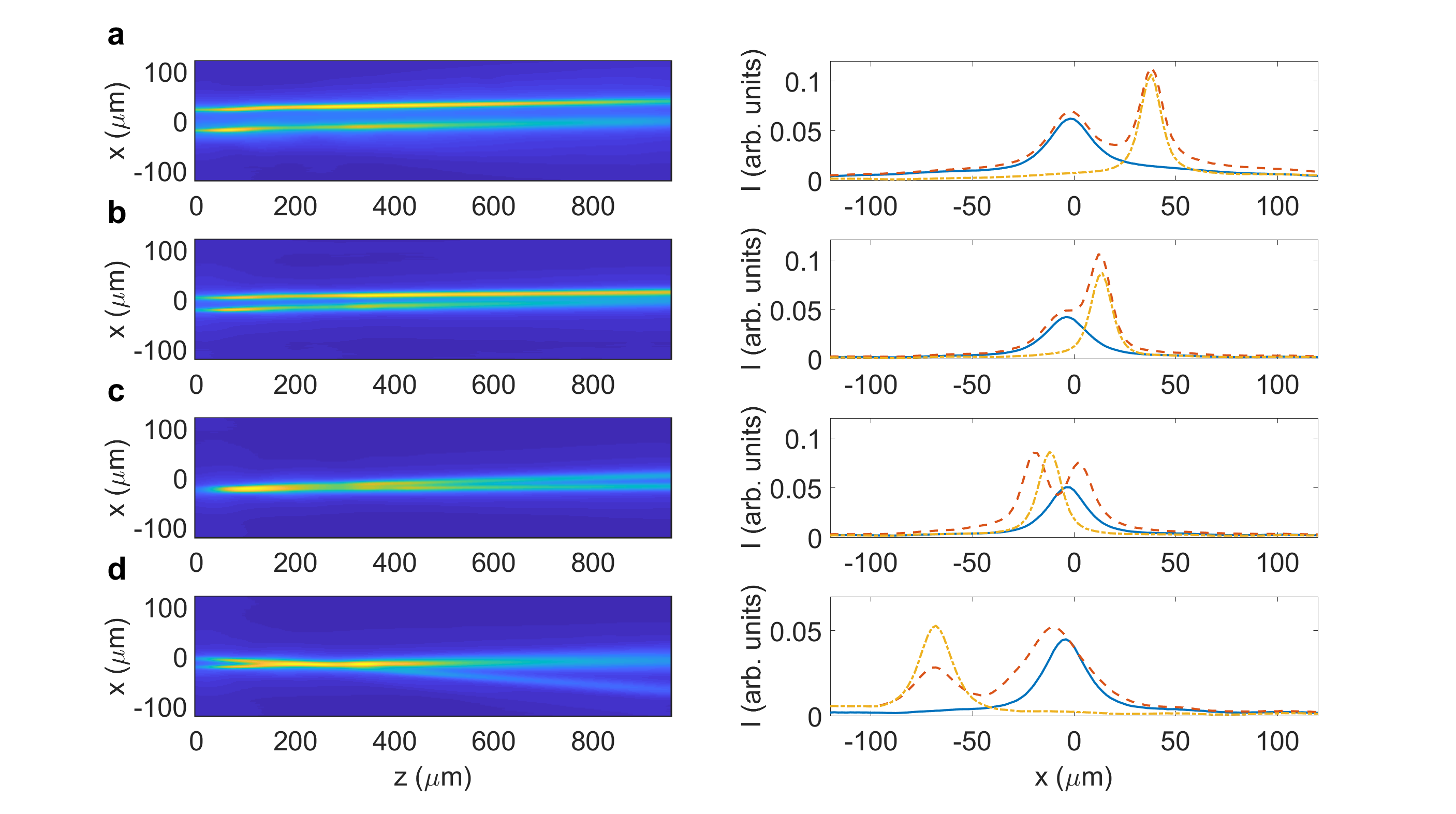}
\caption{\label{fig:two_beams} {\bf Two-beam interaction.} {{\bf a-c}
Interaction of two parallel optical beams separated by a distance of ({\bf a}) $40~\mu$m, ({\bf b}) $20~\mu$m, and ({\bf c}) $0\mu$m. In ({\bf d}) the upper beam is tilted in order to achieve a X-junction configuration. The first column illustrates the intensity distribution on the observation plane $xz$, whereas the second column reports the intensity cross-section at $z=960~\mu$m when the two beam are launched together (orange dashed lines), or when only the upper or the lower beam is launched alone (dot-dashed yellow and blue solid lines, respectively). The input waist is $4~\mu$m, whereas the input powers are 50~mW for the top beam and 40~mW for the bottom beam. }}
\end{figure}

Geometric phase is of primary importance in modern physics \cite{Bhandari:1997,Bliokh:2004,Xia:2009,Dalibard:2011,Sodemann:2015,Lu:2015,Kennedy:2015,Wimmer:2017}.
Geometric phases arise when a classical or quantum system encompasses a variation of its parameters, such as a magnetic dipole moving in a rotating magnetic field \cite{Berry:1984,Aharonov:1987}. Although initially introduced by Berry in the case of cyclic and adiabatic transformation \cite{Berry:1984}, geometric phase has been soon generalized to parameter variations of any kind \cite{Aharonov:1987,Samuel:1988}. 
In optics, geometric phase has been independently discovered by Pancharatnam in 1956 \cite{Pancharatnam:1956} while studying light polarization. Indeed, Pancharatnam found out that photons acquire a phase during polarization evolution, the phase depending on the area subtended by the path on the Poincar\'{e} sphere. This phase is responsible for a new set of optical phenomena, collectively called spin-orbit photonics \cite{Cardano:2015,Bliokh:2015}. The action of the Pancharatnam-Berry phase (PBP) manifests clearly when a circularly polarized beam goes through a half-wave plate: at the output, the helicity is inverted and a phase, equal to twice the rotation angle of the wave plate, is added on the field. This property has been exploited to design and realize a whole new generation of planar and ultra-compact spatial light modulators \cite{Bomzon:2002,Marrucci:2006_1,Arbabi:2015,Maguid:2016,Khorasaninejadeaam:2017,Vella:2018}.
The influence of the Pancharatnam-Berry phase on light propagation in bulk material has been recently investigated \cite{Slussarenko:2016,Alberucci:2016}. It has been demonstrated that a transverse modulation of the geometric phase yields light confinement, despite the absence of any gradient in the refractive index. \\ 
On the other side, the Kerr effect represents one of the cornerstones in nonlinear optics. The Kerr effect consists in a variation in the refractive index $n$ proportional to the beam intensity $I$. In the spatial domain, a positive Kerr effect (i.e., the higher the intensity the higher the refractive index) induces self-focusing \cite{Kelly:1965}. 
For (2+1)D propagation geometries, a pure Kerr effect leads to filamentation and beam collapse \cite{Stegeman:1999}. The collapse can be arrested by using nonlocal materials (the nonlinear perturbation expands beyond the illuminated region) or saturable nonlinearities \cite{Bjorkholm:1974,Suter:1993,Karpierz:1998,Kivshar:2003,Peccianti:2012}. In such cases, a self-trapped beam preserving its shape in propagation -usually called spatial soliton or solitary wave- can be generated \cite{Kivshar:2003}. Regardless of the specific type of nonlinearity, light self-trapping implies a point-dependent change in the refractive index of the material, that is, the formation of light-written waveguides \cite{Stegeman:1999,Duree:1993,Rothschild:2006,DelRe:2010,Peccianti:2012,Fardad:2014,Bezryadina:2017}. Solitons in second-order nonlinear materials, substantially based upon an inhomogeneous generation of second-harmonic, are a notable exception to this general rule \cite{Buryak:2002}. Due to their huge nonlinearity and large degree of nonlocality, in the last few decades light self-trapping has been investigated thoroughly in nematic liquid crystals (NLCs) \cite{Braun:1993,Beeckman:2004,Peccianti:2012,Kwasny:2012}. Essentially, liquid crystals are anisotropic fluids, encompassing properties at an intermediate stage between liquids and solids \cite{Mundoor:2018}. In the nematic phase, liquid crystals behave like inhomogeneous uniaxial crystals. The optical axis is parallel at each point to a molecular field, called director $\hat{n}$ \cite{Degennes:1993}. Consequently, optical waves perceive a refractive index $n_\|$ or $n_\bot$ depending on whether electric fields are parallel or normal to the optical axis, respectively.\\
\begin{figure*}
\includegraphics[width=0.95\textwidth]{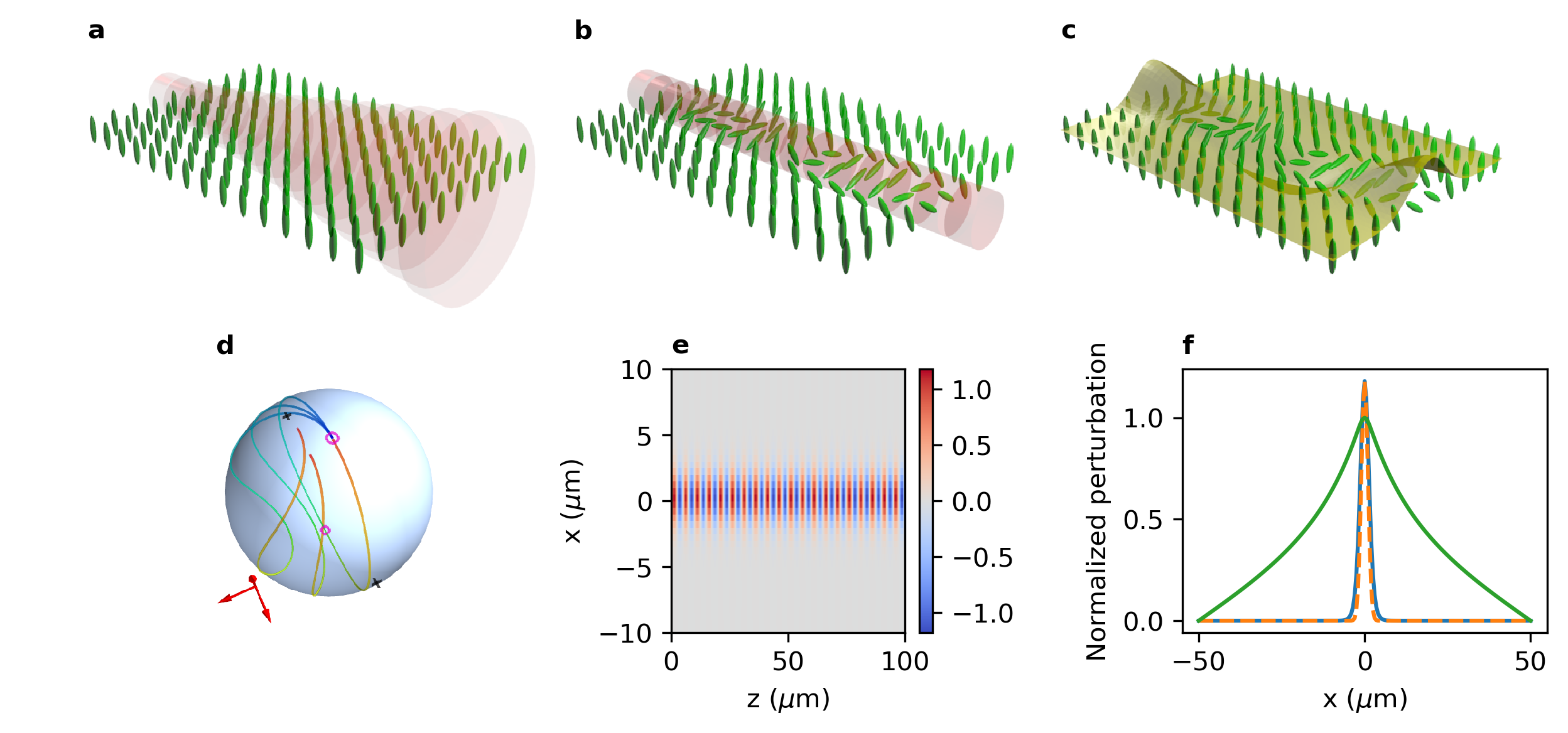}
\caption{{\label{fig:illustration} {\bf Schematic of light propagation through a homeotropically aligned LC material.} {\bf a}, Diffraction of a low power Gaussian beam propagating through the LC medium. {\bf b}, Confinement of a high power Gaussian beam due to the periodic modulation of the LC molecule. The modulation period is $\lambda/\Delta n$. {\bf c}}, In each period the polarization evolves as in a waveplate: The Stokes parameter ${S}_3$ varies from RCP (input) to LCP ($\lambda/ 2\Delta n$) and then back to RCP at the output. {\bf d} Visualization on the Poincar\'{e} sphere of the polarization dynamics of plane waves in one birefringent length.  Angle of the optical axis $\theta$ is supposed to be sinusoidal along the propagation direction. The different lines correspond to a maximum rotation angle $\theta_{max}$ of 5$^\circ$, 15$^\circ$ and 45$^\circ$, respectively (the loop is almost closed for the lowest angles, but it opens up as the angle is increased). {\bf e} Distribution of the rotation $\theta$ (in degrees) on the plane $xz$ and {\bf f} the corresponding cross-section (solid blue line) versus $x$, taken in the section where the rotation angle is maximum; input parameters are $w=2~\mu$m, power density is $2\times$10$^{3}$~Wm$^{-1}$. The dashed line corresponds to the intensity profile $I$, whereas the green solid curve is the light-induced rotation for a uniform (both intensity and polarization) excitation along $z$, i.e. for the case of localization based on dynamic phase.  Here $\lambda=1064~$nm, $n_{\|}=1.7$ and $n_\bot=1.5$. }
\end{figure*}
In this article we will show self-focusing and generation of self-trapped beams based upon a transverse gradient of PBP. As nonlinear mechanism, we will consider reorientational nonlinearity in NLC, the latter consisting in a rotation of the optical axis proportional to the local optical field \cite{Degennes:1993,Peccianti:2012}. We will demonstrate how light is capable to modulate on its own the optical axis both in the transverse plane as well as along the propagation direction, leading to the formation of longitudinally periodic structures with a period $\Lambda=\lambda/\Delta n$, where we defined the wavelength $\lambda$ and the birefringence $\Delta n=n_\|-n_\bot$. In turn, such structures can guide light in spite of the absence of gradient in the refractive index \cite{Slussarenko:2016}, the latter corresponding to the dynamical phase of the system \cite{Berry:1984}. In few words, the strongly coupled light-matter system undergoes a self-adapting process, finally yielding a stable self-trapped propagation over several Rayleigh lengths. \\
The basic mechanism is sketched in Fig.~\ref{fig:illustration}.
We consider a homogeneous distribution of NLC with the director aligned along the $x$ axis (Fig.~\ref{fig:illustration}{\bf a}), illuminated by a circularly polarized (CP) Gaussian beam propagating along the $z$ direction. In the linear regime (i.e., low input powers $P$), the optical torque acting on the molecules is negligibly small: the beam diffracts inside the sample, with the polarization evolving periodically with period $\Lambda$ as in a standard wave plate. Thus, in $z=\Lambda/2$ the beam will be CP but with opposite helicity with respect to $z=0$, whereas it will be linearly polarized along the diagonal and anti-diagonal direction (i.e., forming an angle $\pm 45^\circ$ with the axis $x$) in $z=\Lambda/4$ and $z=3\Lambda/4$. When the input power is increased, molecules undergo an inhomogeneous rotation, i.e., the angle $\theta$ now depends both on the transverse and longitudinal coordinates, respectively (Fig.~\ref{fig:illustration}{\bf b}). Specifically, the rotation $\theta$ will be vanishing when light is CP (i.e., in $z=0$ and $z=\Lambda/2$), whereas it will reach a maximum (minimum) when $z=\Lambda/4$ ($z=3\Lambda/4$), the latter corresponding to the longitudinal positions where the polarization is linear along the diagonal (anti-diagonal) direction. The amplitude of the longitudinal modulation of the optical axis varies across the beam cross-section owing to the inhomogeneous optical torque. Eventually, this corresponds to a transversely-dependent phase modulation due to the net accumulation of PBP \cite{Slussarenko:2016}, in our case leading to the formation of a self-written waveguide. Figure~\ref{fig:illustration}{\bf c} illustrates the corresponding periodic variation in the Stokes parameter ${S}_3={s}_3I=-\frac{\overline{n}}{Z_0}Im(E_xE_y^*)$, where $I=\frac{\overline{n}}{2Z_0}\left(|E_x|^2+ |E_y|^2 \right)$ is the field intensity and $\overline{n}=\frac{n_\bot+n_\|}{2}$ is the average refractive index. To provide a full description of the light polarization, we also introduce the other two normalized Stokes parameters, ${s}_1=\frac{\overline{n}}{2Z_0}\left(|E_x|^2 - |E_y|^2 \right)/I$ and ${s}_2=\frac{\overline{n}}{Z_0}Re(E_xE_y^*)/I$. \\
\begin{figure*}
\includegraphics[width=0.95\textwidth]{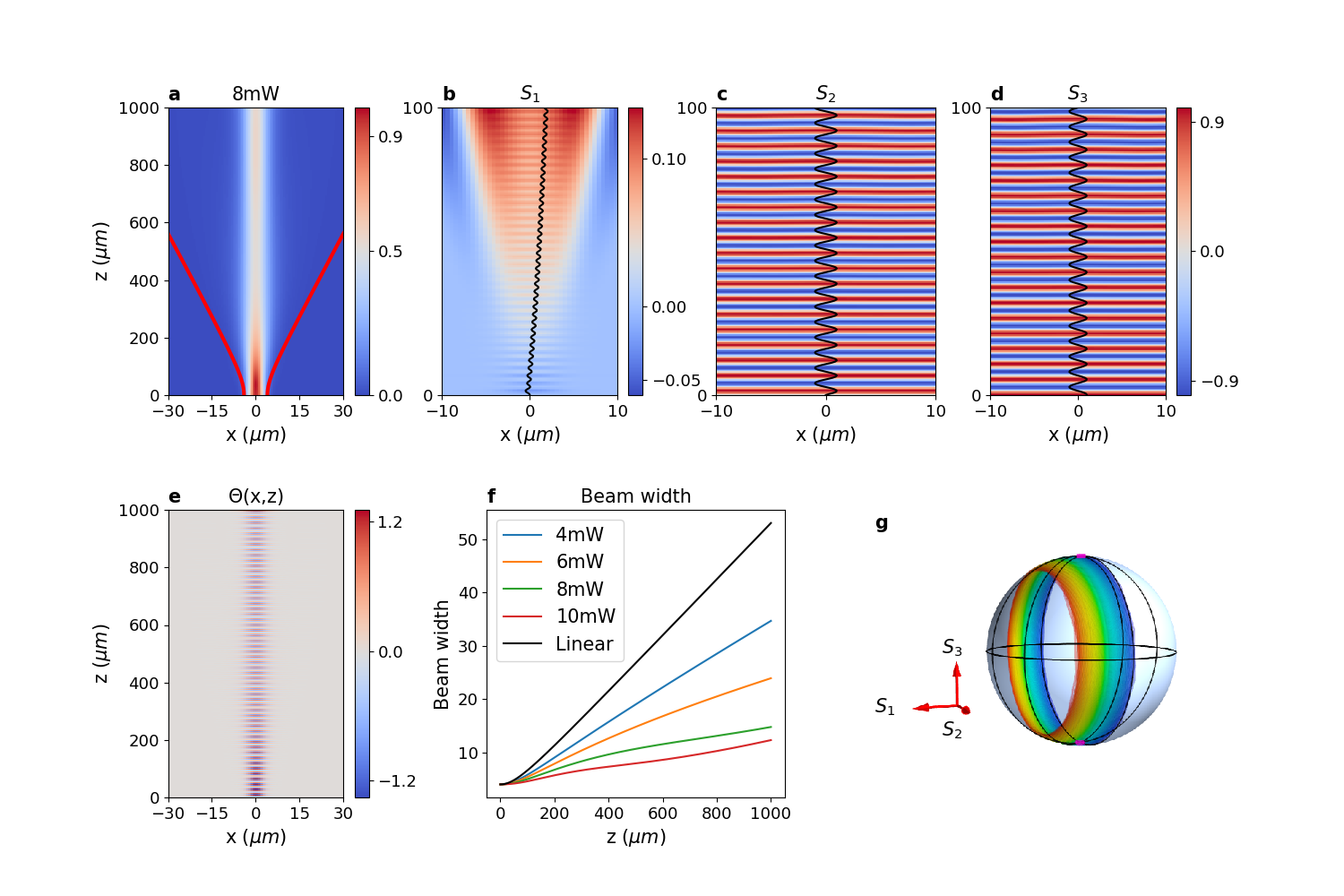}
\caption{{\label{fig:numerical}} {\bf Numerical simulations.} {\bf a}, Evolution of a high power Gaussian beam propagating through a homeotropically aligned NLC cell. Red lines show the linear diffraction. {\bf b-d}, Polarization evolution of the beam, both in the transverse and longitudinal direction, is shown in terms of the Stokes parameters $s_1$, $s_2$ and $s_3$. The black dashed lines show the evolution of the Stokes parameters at the center of the beam. $s_1$ has been magnified $10$ times for a better representation. {\bf e}, Light induced periodic rotation of the optical axis resulting in the confinement of the beam. {\bf f}, Evolution of the width of the beam with decreasing power (bottom to top curve) for a Gaussian beam of input width $w_{in}=4 \mu$m. Topmost curve shows the linear diffraction. {\bf g}, Evolution of the polarization at the center of the Gaussian beam on the Poincar\'{e} sphere. Here the input polarization is RCP, $\lambda=1064~$nm, $n_{\|}=1.7$ and $n_\bot=1.5$. }
\end{figure*}
\begin{figure*}
\includegraphics[width=0.95\textwidth]{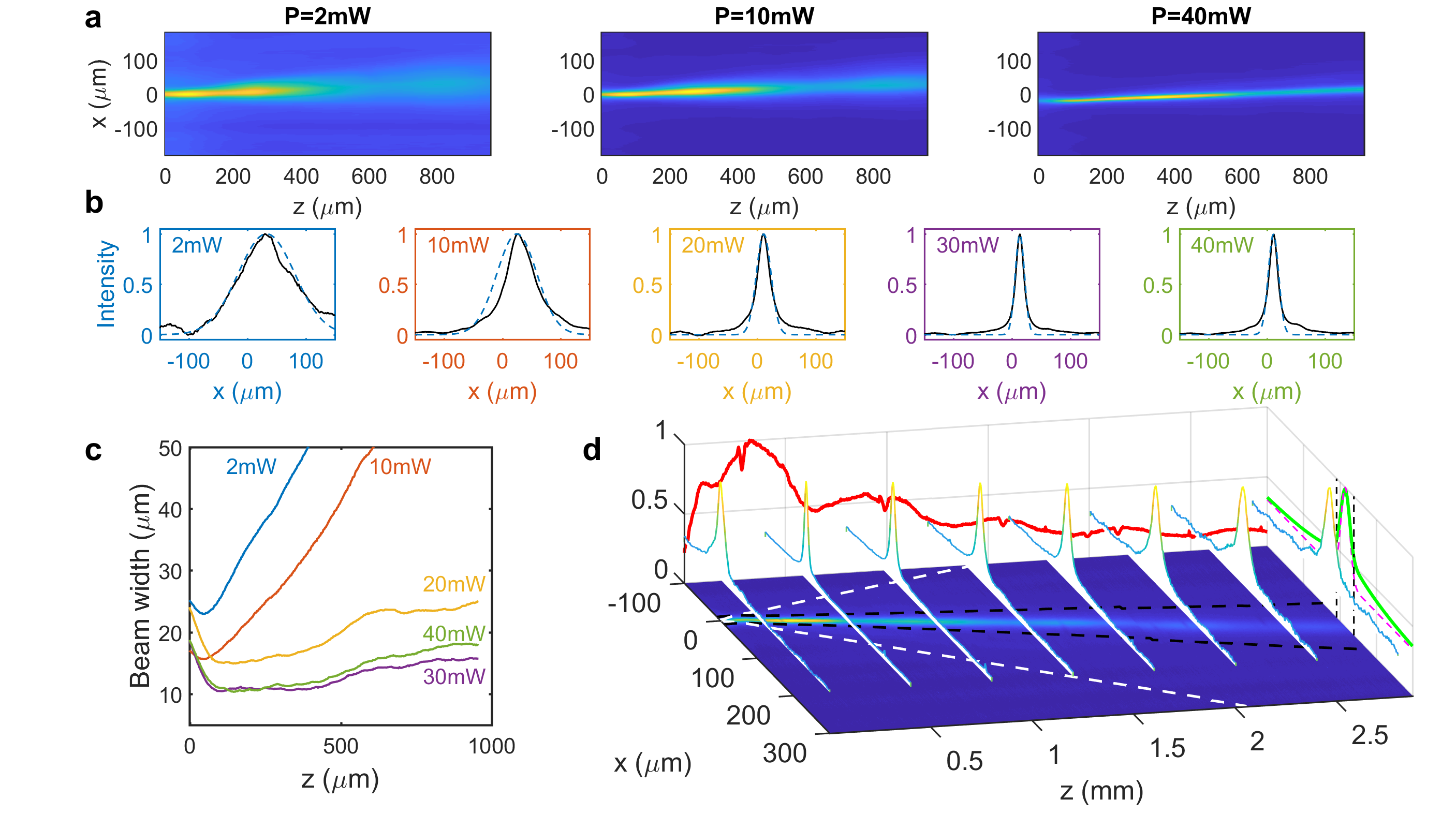}
\caption{{\label{fig:exp_single_beam} {\bf Experimental evidence of light self-localization.} {\bf a}, Light distribution on the plane $xz$ for $P=2~$mW (linear diffraction), 10~mW (mild self-focusing) and 40~mW (self-trapping). {\bf b}, Collection of the beam cross sections at $z=900~\mu$m for several input powers. Black solid and blue dashed lines are the acquired profile and the corresponding best-fitting Gaussian function, respectively.  {\bf c}}, Beam width versus $z$ for the set of powers shown in {\bf b}. {\bf d}, Light evolution when $P=30~$mW, the latter corresponding to the best confinement (see panel {\bf c}). The propagation is shown for over 2.5~mm by collaging three different pictures. The dashed lines is the beam width for Gaussian beams with waists of 2~$\mu$m (white lines) and $10~\mu$m (black lines). The red solid line is the behavior of the intensity peak versus $z$. The green solid (magenta dashed) line is the best-fit to the final beam cross-section, accounting (disregarding) the photon diffusion, respectively. The wavelength is $\lambda=1064~$nm and the input waist is $2~\mu$m, placed in $z=0$. }
\end{figure*}
Let us now reformulate the previous qualitative picture in a more formal way. Writing Maxwell's equations in the stationary regime (electromagnetic field $\propto e^{-i\omega t}$)  in terms of CP waves, the latter being defined as $E_L=(E_x-iE_y)/\sqrt{2}$ (left circular polarization, LCP) and $E_R=(E_x+iE_y)/\sqrt{2}$ (right circular polarization, RCP), light propagation for purely transverse fields obeys
\begin{align}
 \label{eq:MaxwellCP}
  \left(\frac{\nabla^2}{k_0^2} + \overline{\epsilon} \mathbb{I} \right) \left( \begin{array}{c}
                                             E_L \\ E_R
                                            \end{array}\right)&+ \frac{\epsilon_{a}}{2} \left( \begin{array}{cc}
                                             0 & e^{2i\theta} \\ 
																						e^{-2i\theta} & 0
                                            \end{array}\right) \cdot \left( \begin{array}{c}
                                             E_L \\ E_R
                                            \end{array}\right) =0,                                    
 \end{align} 
where we defined $\overline{\epsilon}=\frac{\epsilon_\bot+\epsilon_\|}{2}$,  $\epsilon_a=\epsilon_\|-\epsilon_\bot$ is the optical anisotropy, and $k_0$ is the vacuum wavenumber. Equation~\eqref{eq:MaxwellCP} shows how the light propagation in the paraxial regime is analogous to a massive particle of unitary magnetic moment interacting with a $\theta$-dependent effective magnetic field $\bm{B}_{eff}=\epsilon_a\left[\cos(2\theta)\hat{x}-\sin(2\theta)\hat{y}\right]$ through the coupling energy $\bm{\sigma}\cdot \bm{B}_{eff}$ \cite{Kuratsuj:1998}, where $\sigma_i$ are the three Pauli matrices. Accordingly, the last term in Eq.~\eqref{eq:MaxwellCP} accounts for the spin-orbit interaction related with the anisotropic nature of the material. The spin-orbit interaction is spatially varying through the dependence from the rotation angle\cite{Bhandari:1997} $\theta$,  yielding to an inhomogeneous distribution of geometric phase \cite{Berry:1984}. If $\theta$ is periodically rotated along $z$ (from numerical simulations we can factor out the spatial dependences as $\theta(x,y,z)=H(z)\Gamma(x,y)$, with $H(z+\Lambda)=H(z)$), the PBP accumulates in propagation, and an effective photonic potential  $V(x,y)= -  \frac{2\pi m |\eta_1|}{\Lambda} \Gamma(x,y)$ emerges, where $|\eta_1|$ is the amplitude of the fundamental harmonic of $\theta$ and $m$ is the beam helicity with respect to the propagation distance \cite{Slussarenko:2016}.  \\ 
For plane waves and in the absence of back-reflection, solution of Eq.~\eqref{eq:MaxwellCP} can be found by using the Jones's formalism. Let us make the ansatz $\theta(x,y,z)=\Gamma(x,y)\sin(\frac{2\pi z}{\Lambda})$ in agreement with the periodic variation of polarization in an anisotropic medium. Figure~\ref{fig:illustration}{\bf d} shows on the Poincar\'{e} sphere the polarization evolution from $z=0$ to $z=\Lambda$ for RCP at the input. For small amplitudes $\Gamma$, the polarization path is very similar to the meridian $S_1=0$. When the reorientation increases (i.e., larger input power $P$), the path undergoes strong modifications, and the Stokes parameter $S_1$ differs significantly from zero. \\ 
Equation~\eqref{eq:Maxwell} must\ be solved in conjunction with the reorientational equation, the latter stating the balancing between the elastic (stemming from intermolecular interactions) and the optical torque \cite{Degennes:1993}. In the static ($\partial_t\theta=0$) regime we get 
\begin{align}
   &\frac{1}{\gamma} \nabla^2\theta  \nonumber + \\ &\frac{2Z_0}{\overline{n}} I(x,y,z) \left[ \sin\left(2\theta \right) s_1(x,y,z) + \cos(2\theta) s_2(x,y,z) \right]=0,
    \label{eq:reorientation}
\end{align}
where we have defined the light-matter coupling constant $\gamma=\frac{\epsilon_0 \epsilon_a}{4K}$, being $K$ the Frank's elastic constant \cite{Degennes:1993}. According to Eq.~\eqref{eq:reorientation}, the nonlinear response of the system is nonlocal, that is, $\theta$ changes even in regions where the intensity vanishes \cite{Suter:1993,Conti:2003,Peccianti:2012}. Differently from most of the nonlocal nonlinear media, Eq.~\eqref{eq:reorientation} shows that in LCs the relationship between the field intensity $I$ and the nonlinear director rotation $\theta$ is nonlinear. As a direct consequence, the solution of Eq.~\eqref{eq:reorientation} strongly depends on the input polarization through the Stokes parameters \cite{Santamato:1990}. For example, if the light polarization is normal to the director at rest, linearly polarized inputs move the director only when the impinging power overcomes the Fr\'{e}edericksz threshold \cite{Degennes:1993}. A typical solution of Eq.~\eqref{eq:reorientation} in (1+1)D is shown in Fig.~\ref{fig:illustration}{\bf e}, where we supposed a diffraction-less beam with polarization varying as $s_3=\cos\left(\frac{2\pi}{\Lambda}z\right)$ and $s_2=\sin\left(\frac{2\pi}{\Lambda}z\right)$. As anticipated in our preliminary discussion, the rotation angle $\theta$ follows $s_2$, that is, molecules are periodically rotated by the optical field. Figure~\ref{fig:illustration}{\bf f} compares the cross-section of $\theta$ (blue solid lines) with respect to the optical intensity $I$ (orange dashed lines). The degree of nonlocality is very small with respect to standard reorientational solitons in NLCs, where the nonlocality is fixed by the size of the sample (green solid line) \cite{Rothschild:2006,Peccianti:2012}. The difference is easily explained by noticing that, for small $\theta$, Eq.~\eqref{eq:reorientation} is analogous to a Poisson's equation ruling electrostatics \cite{Rothschild:2006}. As a matter of fact, the periodic forcing term (proportional to $S_2$) corresponds to a vanishing overall charge, thus the long range field along the transverse direction must vanish. \\
We run numerical simulations in (1+1)D of the system composed by Eqs.~(\ref{eq:MaxwellCP}-\ref{eq:reorientation}) to verify our basic idea (see Methods for details). The intensity and the corresponding Stokes parameters distribution are plotted in Fig.~\ref{fig:numerical}{\bf a-d}. As predicted, for large enough powers (in the plotted case the effective power is $\mathcal{P}=8~$mW, see Methods for the definition of $\mathcal{P}$) the optical beam rotates the molecules in a periodic fashion (Fig.~\ref{fig:numerical}{\bf e}), inducing a strong optical self-focusing due to the accumulation of geometric phase in propagation. The larger the input power the stronger the self-focusing is, as shown by the trend of the beam width $w$ versus the propagation distance $z$ and the optical power $\mathcal{P}$ plotted in Fig.~\ref{fig:numerical}{\bf f}. Figure~\ref{fig:numerical}{\bf g} maps on the Poincar\'{e} sphere the polarization in $x=0$ as it evolves along $z$. As predicted above, the Stokes parameter $s_1$ is initially zero, but it increases continuously along the propagation distance, whereas the two remaining Stokes parameters trace periodic orbits on the sphere. Physically speaking, there is a nonlinear exchange of angular momentum between matter and light \cite{Santamato:1990}, the latter counteracting the natural diffraction-induced spreading of light owing to the strong spin-orbit coupling of the system \cite{Marrucci:2006_1,Bliokh:2015}. \\
In order to verify our theoretical predictions, we coupled a circularly-polarized fundamental Gaussian beam at $\lambda=1064~$nm inside a NLC cell of thickness $75~\mu$m (see Methods for details). The intensity distribution on the plane $xz$ can be directly visualized by collecting the scattered light with a customized microscope. Results are plotted in Fig.~\ref{fig:exp_single_beam}{\bf a-b}. For low power, the beam diffracts and free diffraction is observed. As the power is increased ($P=10~$mW), self-focusing is observed, the latter manifesting as a weaker light spreading. At higher powers, diffraction is compensated and a self-collimated beam is excited inside the sample. The behavior of the beam width versus $z$ (Fig.~\ref{fig:exp_single_beam}{\bf c}) is in agreement with the simulations reported in Fig.~\ref{fig:numerical}{\bf f}. Discrepancies at short propagation distances are ascribed to diffusive photons emitted in correspondence to the input interface. The intensity cross-sections encompass exponential tails, the latter increasing in magnitude with $z$ (Fig.~\ref{fig:exp_single_beam}{\bf d}). Such exponential tails are related with the intrinsic time-dependent disorder of NLCs (see Methods). The light self-trapping is evident by comparing the observed propagation with respect to the linear case. The reorientational origin of the phenomena is proved also by the dependence of self-focusing on the direction angle of the linear input polarization: in fact, the maximum self-focusing effect is achieved for input polarizations around the diagonal directions (see Supplementary Material). Figure~\ref{fig:merging} shows the light propagation inside the sample for three different linear polarizations, but now the input wavevector is tilted by $4^\circ$ with respect to the normal at the input interface. The input power is fixed at 50~mW. When the input is vertical (Fig.~\ref{fig:merging}{\bf a}), no appreciable nonlinear effects are observed given that only the extraordinary component is excited in the NLC, hence no optical torque is applied on the molecules. When the polarization is horizontal (Fig.~\ref{fig:merging}{\bf c}) only the ordinary component is propagating inside the sample, the latter undergoing a strong defocusing effect generated at the input interface (see Supplementary Material). Conversely, for diagonal polarizations (Fig.~\ref{fig:merging}{\bf b}) the joint action of the two components yield a substantial molecular rotation in the $xy$ plane, in turn generating a PBP gradient capable to lock the two components and generate a vector self-trapped beam via the spin-orbit coupling.
\begin{figure}
\includegraphics[width=0.49\textwidth]{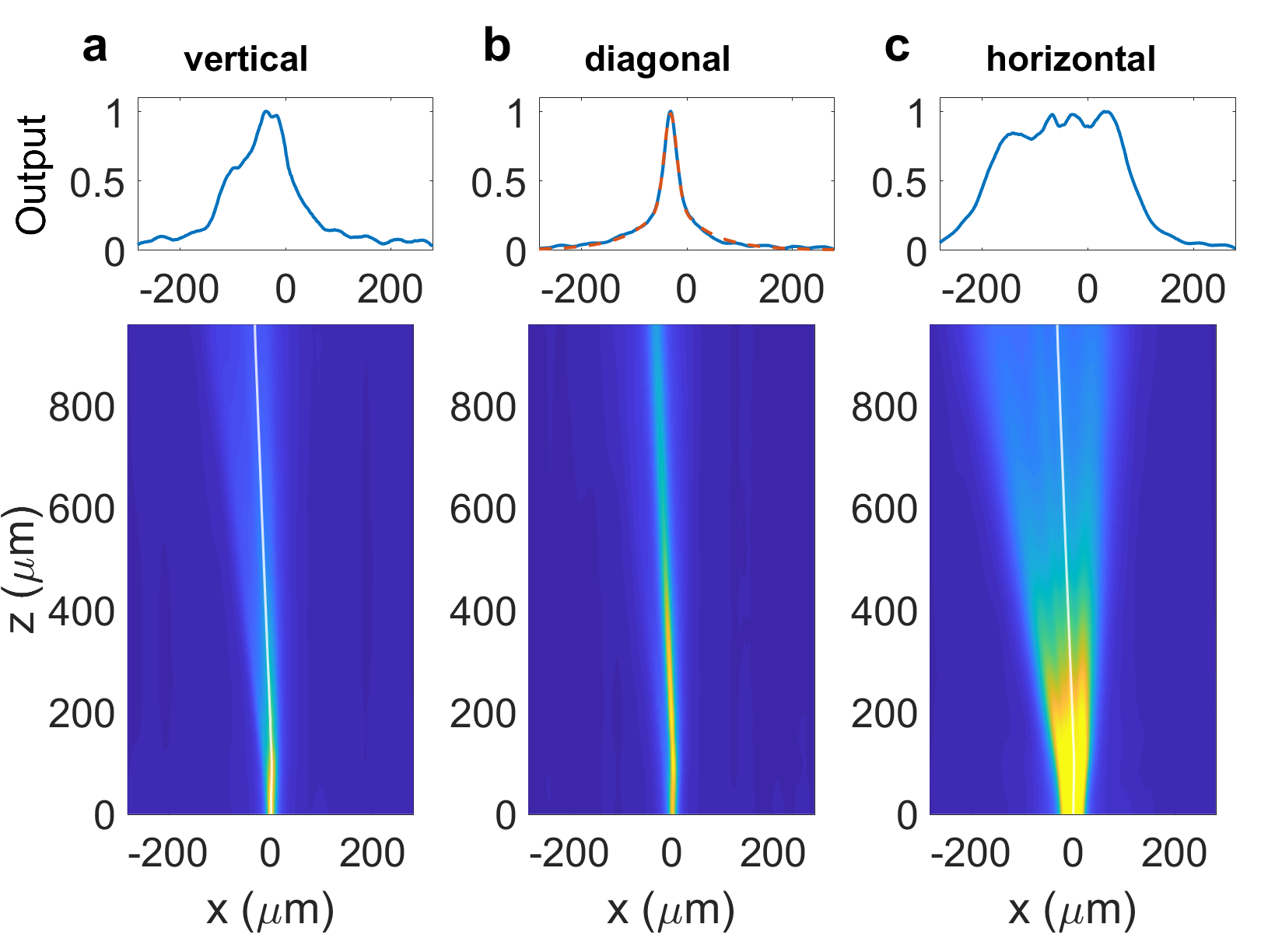}
\caption{\label{fig:merging} {\bf Light behavior for tilted input.} {Light propagation when the input beam is tilted by $\approx 4^\circ$ with respect to the normal at the input interface and is linearly polarized ({\bf a}) parallel to axis $y$, ({\bf b}) at 45$^\circ$, and ({\bf c}) parallel to the axis $x$. Top panels report the intensity cross-section at $z=960\mu$m (solid blue line), whereas the dashed red line is the best-fit accounting for light diffusion (see Methods), providing a width of 20~$\mu$m to be compared with the value of $\approx 50~\mu$m measured in the linear case. The solid white line in ({\bf a,c}) is the trajectory corresponding to the self-trapped beam plotted in ({\bf b}). The input power is 50~mW and the input waist is $4~\mu$m. }}
\end{figure}
Last, we investigated the interaction between two beams encompassing an opposite helicity. Unlike standard self-trapped beams based upon the dynamic phase (see Supplementary Materials), the two beams are expected to repel each other, given that the sign of the photonic potential depends on the sign of the photon helicity at the input. When the relative distance is equal or larger than $20~\mu$m (Fig.~\ref{fig:two_beams}{\bf a-b}), there is not interaction in agreement with the low degree of nonlocality (see the numerical simulations in the Supplementary Materials). When the two fields spatially overlap at the input (Fig.~\ref{fig:two_beams}{\bf c}), the distance between the two beams increases with the propagation distance. 
\\
In conclusion, we demonstrated the self-focusing and self-confinement of electromagnetic waves via the Pancharatnam-Berry phase. The fundamental mechanism is based upon spin-orbit interaction and thus strictly connected to gauge fields \cite{Fang:2012}. Similar phenomena are expected to occur in other systems, such as Bose-Einstein condensates or beams of charged particles under the action of a magnetic field, real or effective. 
With respect to basic physics, our work confirms the relevancy of liquid crystals as a platform for the investigation of angular momentum exchange between light and matter in the nonlinear regime \cite{ElKetara:2012}, including spin-to-orbital conversion and the generation of self-trapped structured beams. The self-localization provides a novel platform to create reconfigurable all-optical devices like directional coupler, beam splitter/combiner and so on based on Pancharatnam-Berry phase and in the absence of refractive index modulation. 
We also envisage the application of these results to the generation of permanent Pancharatnam-Berry waveguides after polymerization of the liquid crystal.

\section{Methods}
\subsection{\bf Light propagation in a twisted liquid crystal.}
At optical frequencies, NLCs are non-magnetic inhomogeneous uniaxial crystals, with the optical axis corresponding to the symmetry axis, named as the director $\hat{n}(\bm{r})$. For fixed external excitations and given boundary conditions, the spatial distribution of $\hat{n}$ in the stationary regime is determined by the balance between all the torques acting on the molecules \cite{Degennes:1993}. In our case, the equilibrium is achieved when the elastic torque balances out the optical one, where the elastic and the optical torques correspond to the first term ($\propto \nabla^2 \theta$) and the second term ($\propto I$) in Eq.~\eqref{eq:reorientation}, respectively. Once the director distribution is known, the optical properties are then determined by the director orientation and by the dielectric eigenvalues $\epsilon_\bot$ and $\epsilon_\|$, the latter being valid for electric fields oscillating normal and parallel to the optical axis, respectively. In fact, the dielectric tensor $\bm{\epsilon}$ is given by $\epsilon_{ij}=\epsilon_\bot+\epsilon_a n_i n_j \delta_{ij}$, where $\epsilon_a=\epsilon_\|-\epsilon_\bot$ is the optical anisotropy. In our theoretical discussion, we consider that the boundary conditions at the cell edges are such that to induce a homeotropic alignment of the molecules along the $y$ axis all around the sample (see Fig.~\ref{fig:illustration}{\bf a}). \\
We consider optical beams impinging normally to the cell, thus having a wavevector parallel to $\hat{z}$. In the paraxial limit, the electromagnetic field lies in the plane $xy$. 
When illuminated, the NLC molecules can rotate only in the plane $xy$ due to the paraxial nature of the field. Thus, the optical properties are fully determined by the rotation angle $\theta$ with respect to the rotation axis $z$. In the stationary regime (electromagnetic field $\propto e^{-i\omega t}$, wavelength $\lambda$ and wavenumber $k_0=2\pi/\lambda$), evolution along $z$ of the electric field $\bm{E}=E_x\hat{x}+E_y\hat{y}$ is ruled by \cite{Jisha:2017} 
\begin{align}
 \label{eq:Maxwell}
  \left(\frac{\nabla^2}{k_0^2} + \epsilon_\bot \mathbb{I} \right) \left( \begin{array}{c}
                                             E_x \\ E_y
                                            \end{array}\right)&+ \epsilon_{a} \left( \begin{array}{cc}
                                             \sin^2\theta & -\frac{\sin2\theta}{2} \\ 
																						-\frac{\sin2\theta}{2} & \cos^2\theta
                                            \end{array}\right) \cdot \left( \begin{array}{c}
                                             E_x \\ E_y
                                            \end{array}\right) =0.                                    
 \end{align}   
Noticeably, there is no transverse refractive index gradient acting on the beam, but only a point-dependent transfer of energy from the $x$ to the $y$-component, i.e., polarization variation, driven by the last term in Eq.~\eqref{eq:Maxwell} \cite{Slussarenko:2016}. Rewriting Eq.~\eqref{eq:Maxwell} in the CP basis, it is straightforward to get Eq~\eqref{eq:MaxwellCP} of the main text.\\
After application of the paraxial approximation \cite{Jisha:2017} and setting $E_{CP}=A e^{ik_0\overline{n}z}$, Equation~\eqref{eq:MaxwellCP} in the main text can be recast as $i\frac{\partial A}{\partial z}= - \frac{1}{2\overline{n}k_0}\left(\frac{\partial^2 A}{\partial x^2} + \frac{\partial^2 A}{\partial y^2}\right) + V(x,y)A$. The effective photonic potential is $ V(x,y)=-\frac{2\pi m |\eta_1|}{\Lambda} \Gamma(x,y) + \frac{1}{4\overline{n}k_0} \left[\left(\frac{2\pi}{\Lambda} \right)^2 \Gamma^2(x,y) + \left(\frac{\partial \Gamma}{\partial x}\right)^2 + \left(\frac{\partial \Gamma}{\partial y}\right)^2  \right]$, where $m=\pm 1$ according to the helicity of the input beam (in our convention $m=1$ corresponds to RCP waves) \cite{Slussarenko:2016}. For maximum rotation angles lower than $45^\circ$ and beam width larger than the wavelength, for sinusoidal modulations along $z$ the photonic potential can be approximated as \cite{Slussarenko:2016} 
\begin{equation}
 V(x,y)\approx -\frac{m\pi}{\Lambda} \Gamma(x,y). \label{eq:phot_potential}
\end{equation}
In the nonlinear case analyzed here, the shape of $\Gamma$ is dictated by the optical beam itself via the reorientation equation~\eqref{eq:reorientation}.

\subsection{\bf Numerical simulations.} 
We simulated the system composed by Eq.~(\ref{eq:MaxwellCP}-\ref{eq:reorientation}) in the (1+1)D limit setting $\partial_y=0$. In the (1+1)D model the initial input field is defined as $E_{in}(x)=\sqrt{\frac{4Z_0 \mathcal{P}}{\pi \overline{n}w_{in}^2}}$, the latter equation thus defining the effective power $\mathcal{P}$ employed in the numerical simulations. As a rule of thumb, the effective power $\mathcal{P}$ is 4-10 times smaller than the real power $P$ in the full 3D case \cite{Alberucci:2010_2}. The electromagnetic equations are solved in the paraxial approximation by using a standard beam propagation method (BPM) encoding operator splitting and Crank-Nicolson algorithm for the diffraction operator.  In solving the optical problem we neglected the losses due to the NLC elastic scattering \cite{Bolis:2017}. In all the simulations we considered the NLC E7 and, accordingly, we used the parameters $n_\bot=1.5$, $n_\|=1.7$ and $K=12\times 10^{-12}~$N, corresponding to NIR radiation at room temperature. The reorientation equation is solved by a Gauss-Seidel algorithm. The two equations are solved jointly via an iterative procedure. First, the optical field is computed with the vectorial BPM for a fixed molecular distribution. The electromagnetic field is then used to compute the new director profile by means of the reorientational equation. The cycle is repeated until convergence is achieved. \\
Our numerical simulations account for the longitudinal nonlocality of the NLC, given that in Eq.~\eqref{eq:reorientation} the second derivative along the propagation direction $z$ is retained. If the term $\partial_z^2\theta$ is neglected, the director distribution would be shaped as in the standard case (see the green curve in Fig.~\ref{fig:illustration}{\bf f}). This is in clear disagreement with the two-beam experiments (Fig.~\ref{fig:two_beams}), where interaction is observed only for quasi-overlapping beams (for a direct comparison with the standard case, see the Supplementary Materials). \\
We finally note that convergence is not achieved for powers beyond a given threshold ($\mathcal{P}_{th}=12$~mW for 1~mm long samples and input waist of 4~$\mu$m), the latter depending on the input parameters, cell size and NLC properties (see Supplementary Materials). Physically, this corresponds to the lack of a stable static state and the appearance of periodic or chaotic oscillations in the NLC, a phenomenon observed also for plane-wave excitations \cite{Santamato:1990}. Accordingly, in the experiments for large powers a quasi-periodic oscillation between two states with different beam widths but same trajectory \cite{Bolis:2017,Alberucci:2018} is observed (see the Supplementary Materials). 

\subsection{\bf Sample preparation and configuration.}

Two float glass substrates (Delta Technologies, CG-90IN) with 1.1~mm thickness are coated with a planar alignment layer (Nylon 6,6) and are consequently rubbed with a velvet cloth to ensure a certain orientation (in our case parallel to the axis $z$) of the LC at the boundary. On the two glasses an ITO layer is deposited to permit the application of a voltage to reorient the LC molecules. The two substrates are glued together by depositing a glue pattern onto one of the substrates.  Anti-parallel alignment is used. Spherical spacer balls (with thickness 75~$\mu$m) are dispersed in the glue before applying the glue onto the substrate. The two substrates are assembled together and the glue is consequently cured using UV light.\\
A third glass substrate with 100~$\mu$m thickness is coated with a homeotropic alignment layer (SE-1211). This substrate is fixed perpendicularly onto the sides of the two assembled substrates using the same UV curable glues (with spacers of 10~$\mu$m). All the remaining edges of the assembled device are carefully closed using UV curable glue. Only two holes are left open through which the liquid crystal (E7, Merck) is filled in by using capillarity. These openings are finally also closed using a drop of 2-component epoxy glue.\\
To induce vertical orientation of the director in the cell bulk, a sinusoidal voltage at 9~V (peak value, frequency 1~kHz) is applied to the cell. Noteworthy, the quasi-static field induces a director rotation in the plane $yz$; accordingly, we introduce the angle $\varphi$ between the director and the propagation axis $z$. The reorientation angle in the  absence of light illumination can be computed exactly by jointly solving the LC reorientational equations and the Poisson equation for the quasi-static electric field. For symmetry reasons, the maximum is placed in the cell mid-plane with respect to the vertical axis $y$. For the given bias, a maximum rotation angle $\varphi_m=$89.9$^\circ$ is found (the governing equations and related numerical computations are reported in the Supplementary Materials). To determine the role played by the dynamic phase in our experiments, we need to assess the maximum jump in the extraordinary refractive index (defined in this case by $n_e=\left( \frac{\cos^2 \varphi}{\epsilon_\bot}+ \frac{\sin^2 \varphi}{\epsilon_\|}\right)^{-1/2}$) available in the case of light-induced changes in the angle $\varphi$ \cite{Peccianti:2012}. To do that, we need to compute the difference between $n_e(\varphi=\pi/2)=n_\|$ and $n_e(\varphi_m)$. Such difference in our case is $\approx 6\times 10^{-7}$, which is clearly negligible. In fact, in the case of an index well of size $75~\mu$m, the corresponding nonlinear mode would be $\approx 3~$mm wide, much larger than what we observed in the experiments. Finally, as shown in the Supplementary Materials, the rotation angle $\varphi$ along $y$ (in the $x$ direction the director is homogeneously distributed) is uniform in an interval of $\pm 20~\mu$m around the center of cell (within an accuracy of $1^\circ$).  

\subsection{\bf Experimental setup.}
In the experiments we employed a CW laser at 1064~nm with a coherence length of 4~mm. In the single beam case, the laser radiation is coupled into the NLC planar cell by using a 10X microscope objective (MO) (providing a waist of about 4~$\mu$m) or an aspheric lens Thorlabs A240-TM (providing a waist of about 2~$\mu$m). The beam waist is placed at the input interface to facilitate the observation of self-focusing and self-trapping. A cascade HWP (half wave-plate)-polarizer-QWP is used to control both the power and the polarization of the beam injected into the sample. For the experiments concerning linear input polarizations, the QWP is exchanged with another HWP mounted on a rotation stage. \\
For the two-beam experiment, the beam is divided equally by a polarizing beam splitter (PBS). The path of one of the two paths contains a metallic mirror mounted upon a translational stage. The two beams are then recombined by using a second identical PBS. By proper combination of rotation and translation of the mirror, the position and the direction of the second beam into the NLC cell can be adjusted with respect to the reference one. Due to the PBS, one beam is vertically polarized, whereas the second beam is horizontally polarized. When a QWP is placed after the second PBS, two circularly polarized beams of opposite helicities are excited. 
Finally, the light distribution over the plane $xz$ into the NLC can be visualized by collecting with a 2X microscope objective the light scattered by the sample along the $y$ direction \cite{Peccianti:2012}. A sketch of the experimental setup is provided in the Supplementary Materials.

\subsection{\bf Characterization of the beam properties inside the NLC sample.}
The experimental profiles plotted in the main text are averaged over 100 samples to smooth out the time-dependent noise arising from the NLC scattering \cite{Bolis:2017}. The averaged intensity distribution is then filtered with a Gaussian filter. The local beam width and position are found by a best-fitting procedure. In this paper we chose two types of functions: a Gaussian (in the form $I_G=I_0e^{-2(x-\overline{x})^2/w^2}$) and its convolution with a double exponential function (i.e., a Laplace distribution $R(x)=R_0e^{-|x|/l}$). The convolution $F(x)=I_G(x)*R(x)$ can be expressed in closed form as
\begin{align}
 & F(x)=\sqrt{\frac{\pi}{2}}\frac{I_0 R_0 w}{2}e^{\frac{w^2}{8l^2}} \Biggl\{ e^{-\frac{x}{l}} \text{erfc} \left[\frac{\sqrt{2}}{w} \left(\frac{w^2}{4l}-x\right)  \right] \nonumber  \\ &+ e^{\frac{x}{l}}\text{erfc} \left[\frac{\sqrt{2}}{w} \left(x+\frac{w^2}{4l}\right) \right] \Biggr\}.
\end{align}
For the Gaussian case, the two varying parameters are the beam position $\overline{x}$ and the beam width $w$. When the convolution is employed the overall fitting function is $I_G(x)+F(x)$. In the latter case there are two additional free parameters: the maximum intensity $I_0$ of the unscattered field, the diffusion length $l$, and the scattering intensity $R_0$. The function $F$ is chosen to account for the multiple scattering occurring in NLC, the latter responsible for the emergence of exponential tails as finite-size beams propagate inside the material (see Ref.~\cite{Alberucci:2018} and citations therein).

\newpage

\bibliographystyle{naturemag}
\bibliography{references}

\vspace{1 EM}

\noindent\textbf{Acknowledgments}

\noindent We thank Frederik Vanacker for his help in the laboratory and Raouf Barboza for critical reading of the manuscript. A.A. is supported by Deutsche Forschungsgemeinschaft (DFG) via the International Research Training Group GRK 2101.  

\vspace{1 EM}

\noindent\textbf{Author Contributions}

\noindent A.A. and C.P.J. conceived the work, developed the theory, designed and carried out the experiments, performed the numerical simulations and wrote the paper; J.B. prepared the samples; all authors discussed the results and contributed to the manuscript.

\vspace{1 EM}

\end{document}